\documentclass{amsproc}

\theoremstyle{definition}

\theoremstyle{remark}

\numberwithin{equation}{section}

\usepackage{color}

\begin{document}

\title{Spin Foam State Sums and Chern-Simons Theory}

\author{Aleksandar Mikovi\'c}
\address{Departamento of Matem\'atica, Faculdade de Engenharias e Ci\^encias Naturais, Universidade Lus\'ofona, 1749-024 Lisboa, Portugal}
\email{amikovic@ulusofona.pt}
\thanks{The first author was supported in part by the FCT grant PTDC/MAT/2006 and by Mathematical Physics Group of the University of Lisbon.}

\author{Jo\~ao Faria Martins}
\address{Departmento de Matem\'atica, Faculdade de Ci\^{e}ncias e Tecnologia, Universidade Nova de Lisboa, 2829-516 Caparica, Portugal}
\email{jn.martins@fct.unl.pt}
\thanks{The second author was supported  by CMA, through Financiamento Base 2009 ISFL-1-297 from FCT/MCTES/PT
}

\date{January 30, 2010, in revised form ???}



\begin{abstract}
We review the spin foam state-sum invariants of 3-manifolds, and explain their relationship to manifold invariants coming from the Chern-Simons theory. We also explain the relationship 
between the known invariants of spin networks by using the Chain-Mail formalism of J. Roberts. This formalism can be understood as a quantum-group regularization of the BF theory path integrals.
\end{abstract}

\maketitle

\section{Spin Foams}

Spin foams are colored two-complexes and they naturally appeared in the context of Loop Quantum Gravity, see \cite{B}. They can be viewed as world-sheets of spin networks propagating in time, where the spin networks are colored graphs, originally invented by Penrose \cite{P}, representing the states of quantum geometry on a three-dimensional space.  One can then imagine a spin foam interpolating between two fixed spin networks, so that by associating a complex number to the spin foam, which depends on the spin foam labels, and then summing over the labels, one can then obtain an invariant of the spin networks, provided that the sum over the labels is independent of the triangulation of the four manifold whose dual two-complex is the given spin foam. This sum over the labels is called the spin foam state sum, and the rules how to find triangulation independent spin foam amplitudes can be inferred from the path integrals for BF theories, see \cite{B}.

A $n$-dimensional BF theory can be defined by the action
\begin{equation} S = \int_M Tr\,( B \wedge F ) \,,\end{equation}
where $M$ is a $n$-dimensional compact manifold, $F = dA + A\wedge A$ is the curvature 2-form for a connection $A$ on the principal bundle $P_G (M)$ for a compact Lie group $G$, $B$ is a $\bf g$-valued $(n-2)$-form, where $\bf g$ is the Lie algebra  of $G$. The path integral for this BF theory can be written as a functional integral
\begin{equation} Z(M) =\int DA \exp\left(i\int_M Tr\,( B \wedge F ) \right) \,,\end{equation}
which can be formally written as a functional integral over the flat connections on $M$
\begin{equation} Z(M) = \int DA \,\delta (F)\,,\label{fcpi}\end{equation}
where $\delta (F)$ denotes the product $\prod_{x\in M} \delta (F(x))$ and $\delta (x)$ is the Dirac delta-function. 

All these expresions are formal and have to be defined, and the standard way is through a triangulation $\Delta(M)$ of $M$. The idea is that $Z(M)$ should be a manifold invariant, and since the BF theory is a topological theory, then a discretization of the path integral (\ref{fcpi}) will not depend on a chosen triangulation of $M$. Let us replace $\Delta(M)$ with its dual simplical complex $\Delta^* (M)$ and consider a discretization of (\ref{fcpi}) given by
\begin{equation} Z(M) = \int \prod_l dg_l \prod_f  \,\delta (g_f)\,,\label{dpi}\end{equation}
where $l$ are the dual edges and $f$ are the dual faces, $g_l$ is the holonomy along an edge $l$ and $g_f = \prod_{l\in\partial f} g_l$ is the holonomy along the boundary loop $\partial f$ of $f$.

The group integrations in (\ref{dpi}) can be performed by using the Peter-Weyl theorem
\begin{equation} \delta (g) = \sum_{\Lambda} \,\textrm{dim}\Lambda \, Tr\left(D^{(\Lambda)}(g)\right)\,, \end{equation}
and
\begin{equation} \int_G dg \, D^{(\Lambda_1)}(g)\otimes\cdots \otimes D^{(\Lambda_m)}(g) = \sum_{\iota}\,C^{(\iota)}\otimes\overline{C^{(\iota)}}\,, \label{gii}\end{equation}
where $D^{(\Lambda)}(g)$ is the matrix of the group element $g$ in the irreducible representation $\Lambda$ and $C^{(\iota)}$ is the tensor for the intertwiner $\iota$ from the tensor product of representations $\Lambda_1 , \cdots, \Lambda_m$. Then
\begin{equation} Z(M) = \sum_{\Lambda,\iota}\prod_f \textrm{dim}\Lambda_f \, \prod_v A_v (\Lambda , \iota )\,,\label{sfss}\end{equation}
where the sum is over the colorings  $\Lambda = (\Lambda_1 , \cdots, \Lambda_N )$ of the faces of the dual simplical complex by the irreps of $G$. The sumation has also to include the colorings $\iota = (\iota_1 , \cdots, \iota_M )$ of the dual edges by the corresponding intertwiners, and $A_v$ is the evaluation of the $n$-simplex spin network in terms of the products of $C^{(\iota)}$. 

In the case when $G=SU(2)$, the irreps are called spins, and hence the name ``spin foam'' for the colored dual two-complex. For a three-dimensional manifold, the dual faces of a triangulation correspond to the edges, and the dual vertices correspond to the tetrahedrons, so that $A_v$ is the $6j$ symbol.

The spin-foam state sum (\ref{sfss}) can be shown to be independent of the triangulation, due to the invariance under the $n$-dimensional Pachner moves. However, the sum (\ref{sfss}) is infinite and therefore divergent in general case, so that one still does not have a well-defined invariant. Fortunately, the formal expression (\ref{sfss}) can be regularized by replacing the irreps of $G$ by the irreps of the quantum group $U_q ({\bf g})$, where $q$ is a root of unity. This can be done because there is a one-to-one correspondence between the irreps of a compact Lie group and the irreps of the corresponding quantum group when $q$ is not a root of unity. When $q$ becomes a root of unity, this correspondence still holds for the irreps of non-zero quantum dimension. Consequently, all the spin network evaluations, which are the simplex weights in (\ref{sfss}), can be replaced by quantum spin network evaluations. The invariance under the Pachner moves is maintained and (\ref{sfss}) becomes a finite sum, because when $q$ is a root of unity there are only finitely many irreps $\Lambda$ of non-zero quantum dimension.

Therefore
\begin{equation} Z(M) = \kappa^{N}\sum_{\Lambda,\iota}\prod_f \textrm{dim}_q \Lambda_f \,\prod_l A_l^{(q)}(\Lambda,\iota) \prod_v A_v^{(q)} (\Lambda , \iota )\,,\label{qsf}\end{equation}
is a manifold invariant, where the label $q$ denotes the quantum group evaluations of the relevant spin networks. The factor $\kappa$ is related to $q$, $N$ is the number of $0$-simplices in the triangulation and a dual-edge amplitude $A_l$ appears\footnote{One can set $A_l =1$ by changing the normalization of the spin network evaluations.}. In the three-manifold case and $G=SU(2)$ this is the Turaev-Viro invariant \cite{TV}, while in the 4-manifold case (\ref{qsf}) gives the Crane-Yetter invariant \cite{CY}.

\section{Chern-Simons and BF theory path integrals}

A BF theory is a theory of flat connections on a manifold, and in three dimensions one also has the Chern-Simons (CS) theory, which is also a topological theory of flat connections. Given that the spin-foam state sum invariants are derived from the BF theory, the question is what is the relation to the manifold invariants derived from the CS theory? Witten showed that the CS invariants can be derived from the CS path integral \cite{W}, and the path-integral approach can be used to find the relation between the spin foam and CS invariants.

A BF theory with a cubic interaction is given by the action
\begin{equation} S_{\lambda}(A,B) = \int_M Tr\,( B \wedge F + \lambda\, B\wedge B\wedge B ) \,,\end{equation}
where $\lambda$ is a parameter. In the $SU(2)$ group case, this action corresponds to the
three-dimensional (3d) General Relativity (GR) theory with a cosmological constant $\lambda$. It is easy to show that
\begin{equation} S_{\lambda}(A,B) = S_{CS}(A_+) - S_{CS}(A_-)  \,,\label{spm}\end{equation}
where $A_\pm = A \pm \frac{B}{\sqrt\lambda}$, $S_{CS}(A)$ is the Chern-Simons action for a connection $A$
\begin{equation} S_{CS}(A) = \frac{k}{2\pi}\int_M \langle A \wedge dA + \frac{2}{3} A\wedge A\wedge A \rangle \,,\end{equation}
and $k=2\pi/\sqrt\lambda$. Witten has showed that the CS action is well-defined if $k$ is an integer \cite{W}, and this integer is related to the order of the root of unity taken for $q$ in the quantum group $U_q (\bf g )$. In the $SU(2)$ case, $q=\exp(\frac{\pi i}{k+2})$.

By using (\ref{spm}) one can formally write
\begin{equation} \int DA \,DB e^{ iS_{\lambda}(A,B)}=\int DA_+ DA_- e^{iS_{CS}(A_+) - iS_{CS}(A_-)}=|Z_{CS}(M)|^2 \,,\label{bfcs}\end{equation}
where $Z_{CS}(M)$ is the Chern-Simons path integral. Witten, Turaev and Reshetikhin have shown that this path integral can be rigorously defined \cite{W,RT}, and the corresponding manifold invariant is proportional to the quantum group evaluation of a surgery link of $M$. We denote this invariant as $Z_{WRT}(M)$.

On the other hand, the Turaev-Walker theorem \cite{T} implies
\begin{equation} Z_{TV} (M) = |Z_{WRT}(M)|^2 \,,\label{twt}\end{equation}
where $Z_{TV}(M)$ is the spin foam state-sum invariant (\ref{sfss}) for the 3-manifold $M$. From (\ref{twt}) and (\ref{bfcs}) it follows that $Z_{TV}$ is the path integral for the three-dimensional euclidean General Relativity theory with a cosmological constant. 

Note that the parameter $\lambda$ can take only the restricted values $4\pi^2 /k^2$, $k\in \bf N$ in the case of the path integral $Z(\lambda,M)$ for 3d GR. A natural question is how to define this path integral for the case when $\lambda \ne 4\pi^2 /k^2$. This can be done by using the spin foam perturbation theory \cite{M2}, and the result can be again expressed in terms of the $Z_{WRT}(M)$ invariant. Namely
$$ Z(\lambda, M) = Z_{TV}(k,M)\big{|}_{k=\frac{2\pi}{\sqrt\lambda}}\,,$$
where the right-hand side of the equation represents the function $Z_{TV}(k,M)=|Z_{WRT}(k,M)|^2$ evaluated at a non-integer point $k=2\pi/\sqrt\lambda$.

\section{Spin network invariants}

Witten showed in \cite{W} that the Jones polynomial $J_K (q)$ of a knot $K$ embedded in a three-sphere $S^3$ can be obtained by analysing the following path-integral invariant of $K$ 
\begin{equation} \langle K \rangle = \int DA e^{iS_{CS}(A)}W_K (A) \,,\label{kwl}\end{equation}
where $W_K(A)$ is the trace of the holonomy of a connection $A$ around $K$. The invariant $\langle K \rangle$ depends on the representation of the $SU(2)$ group used to define the holonomy. When $K$ is an unknot, then
$$ \langle K \rangle = \textrm{dim}_q j =(-1)^{2j}\, {q^{2j+1} - q^{-2j-1}\over q - q^{-1}}\,,$$
where $j$ is the spin of the $SU(2)$ representation and the Jones polynomial corresponds to $j=\frac{1}{2}$.

Beside colored loops, one can consider more general objects, like spin networks $\Gamma$, which are oriented graphs whose edges are colored by the spins. In analogy to (\ref{kwl}), one can consider a spin-network invariant
\begin{equation} \langle \Gamma \rangle = \int DA \,e^{iS_{CS}(A)}W_\Gamma (A) \,,\label{wlcs}\end{equation}
where $W_\Gamma (A)$ is given by the product of the edge holonomies contracted by the intertwiner tensors $C^{(\iota)}$
$$ W_\Gamma (A) = \prod_{l\in E(\gamma)} \textrm{Hol}_{l} (A) \prod_{v\in V(\gamma)} C^{(\iota_{v})}\,,$$
where $E(\gamma)$ and $V(\gamma)$ are the sets of the edges and the vertices of the graph $\gamma$ corresponding to the spin network $\Gamma$.

One can be even more general, and consider that $\Gamma$ is embedded in an arbitrary closed 3-manifold $M$. In any case, it can be shown that the path-integral (\ref{wlcs}) can be rigorously defined as the Witten-Reshetikhin-Turaev invariant
$Z_{WRT} (M,\Gamma)$, which is proportional to  the quantum group evaluation of the link $L_M \cup \gamma$ embedded in $S^3$, where $L_M$ is any surgery link of $M$ \cite{W,RT}.

Given that (\ref{wlcs}) can be rigorously defined in CS theory, a natural question is how to define the analogous invariant in the BF theory case. One way is to start from the path-integral expression
$$ \langle \Gamma \rangle_{BF} = \int DA DB \,e^{i\int_M \langle B\wedge F \rangle} W_\Gamma (A)\,,$$
and to use the spin-foam discretization to define it. By repeating the procedure from section 1 for $\Gamma$ embedded in the handle-body $H$ obtained from thickening the dual one-complex of the triangulation of $M$, where each edge of $\Gamma$ goes along some one-handle of $H$ and each  vertex of $\Gamma$ is placed in some zero-handle of $H$, then a Turaev-Viro type state sum is obtained. In the $SU(2)$ case this sum takes the form \cite{M,M1}
\begin{equation} Z_{TV}(\Gamma,M) = \sum_{J,\iota}\prod_f \textrm{dim}_q j_f \, \prod_{v\notin V_{\gamma}} \{6j\}_v^{(q)} \prod_{v\in V_{\gamma}} A_v^{(q)} (j ,j_\Gamma, \iota,\iota_\Gamma )\,,\label{sfmi}\end{equation}
where $V_\gamma$ are the vertices of the dual 1-complex whose thickenings contain the vertices of the graph $\gamma$ of $\Gamma$. $A_v^{(q)} (j ,j_\Gamma, \iota,\iota_\Gamma)$ is the quantum group evaluation of a tetrahedron spin network with an insertion corresponding to the spin network vertex residing in the 3-handle of $v$, where $\{j_\Gamma , \iota_\Gamma \}$ are the spins and intertwiners of $\Gamma$.

By using the Chain-Mail formalism of J. Roberts \cite{R}, it can be shown that $Z_{TV}(\Gamma,M)$ can be represented as the quantum group evaluation of a link embedded in $S^3$ \cite{M1}, and hence a relationship to the invariant $Z_{WRT}(\Gamma,M)$ can be obtained. The reason is that a Chain-Mail link $CHL_M$ for $M$ can be constructed by using the handle-body $H$, which was a thickening of the dual one-complex $\Delta_1^* (M)$ of a triangulation of $M$. Let $N_p$ be the number of $p$-handles of $H$, then the link $CHL_M$ consists of $N_2$ loops runing along the one-handles of H, which correspond to the boundaries of the faces of $\Delta_1^* (M)$ and $N_1$ loops running around the middle sections of the one-handles of $H$.  

Then it can be shown for $G=SU(2)$ and a trivalent spin network\footnote{In the $SU(2)$ group case, a spin network vertex of valence $m > 3$ requires an intertwiner. However, such a vertex can be represented as a tri-valent spin network with $m-2$ vertices.} that
\begin{equation} Z_{TV}(\Gamma,M) = \eta^{-N_0 - N_2}\langle CHL_M \cup \gamma, \Omega^{N_1 + N_2} , j_\Gamma \rangle \,,\label{chm}\end{equation}
where 
$$ \eta^2 = \sum_{j=0}^{k/2}(\dim_q j )^2 = {k+2\over \sin^2\left( \frac{\pi}{k+2} \right)}$$
and $\Omega$ is the coloring of each loop of $CHL_M$. The $\Omega$-element is a linear combination of all possible colorings of a loop, and it is given by the sum
$$\Omega = \sum_{j=0}^{k/2} (\dim_q j )\, C(j) \,,$$ 
where $C(j)$ denotes the coloring of a loop by spin $j$. 

The derivation of (\ref{chm}) is the same as the derivation of the Lie group state sum (\ref{sfss}) from the group integral (\ref{dpi}), since both derivations are based on the same graphical calculus, which is associated to the equation (\ref{gii}), see \cite{B,M1}.

Given that the pair $(CHL_M,\Gamma)$ is a  surgery link  for  the  spin-network $\Gamma$ embedded in the connected sum $M\# \overline{M}$, where $\overline{M}$ is the manifold $M$ with the opposite orientation and $\Gamma$ is embedded in the obvious way in the $M$ part of $M\# \overline{M}$, it follows from (\ref{chm}) that
\begin{equation} Z_{TV}(\Gamma,M) = Z_{WRT}(\Gamma,M)\overline{Z_{WRT}(M)} \,.\label{tvw}\end{equation}

The formula (\ref{tvw}) has several interesting consequences:
\begin{itemize}
\item It gives a state-sum representation of $Z_{WRT}(\Gamma,M)$, since
$$Z_{WRT}(\Gamma,M) = \left(\overline{Z_{WRT}(M)}\right)^{-1} Z_{TV}(\Gamma,M)\,.$$

\item It allows for a path-integral interpretation, which is
$$Z_{TV}(\Gamma,M) = \int DA \,DB e^{ iS_{\lambda}(A,B)}W_\Gamma (A_+)\,, $$
where $\lambda = 4\pi^2 /k^2$.

\item It gives the relation to Turaev's shaddow-world invariants, see \cite{T}. Namely, Turaev has defined a state-sum invariant of a spin network $\Gamma$ embedded in $M$, which is based on the $Z_{TV} (M)$ sum in the following way. Let $M'$ be a 3-manifold obtained by excising a tubular neighborhood of $\gamma$ from $M$, where $\gamma$ is the graph of $\Gamma$. Triangulate $M'$ and construct the corresponding $Z_{TV} (M')$ such that the $6j$ symbols for the terahedrons which have a triangle on the triangulation of the boundary surface $\Sigma$ of $M'$ are excluded. Instead of the boundary tetrahedrons weights, one includes in the state sum the weight associated with a graph $\gamma'\cup \delta$ where $\gamma'$ is a natural projection of $\gamma$ onto $\Sigma$ and $\delta$ is the dual one-simplex of the triangulation of $\Sigma$ induced by the triangulation of $M'$.

The graph $\gamma'\cup\delta$ divides the surface $\Sigma$ into disjoint discs, which can be colored by spins $s$. One also has the spins associated to the edges of $\gamma'$, which are the spins of the spin network $\Gamma$, and the spins associated to the edges of $\delta$, which are the spins of a coloring $J$ of $\Delta (M')$. There is a natural way to associate the weights to such a coloring of $\Sigma$ in terms of the quantum $6j$ symbols, see \cite{T,M1}. By making the product of the surface weights and summing over the surface spins $s$ one obtains the weight $w_\Sigma (\Gamma,j)$ so that
$$ \tilde{Z}_{TV} (\Gamma, M) = \sum_{J}\prod_{j\in J} \dim_q j \prod_{v\in M'} \{6j\}_v \,w_\Sigma (\Gamma , j )\,.$$

Turaev has shown that $\tilde{Z}_{TV} (\Gamma, M)$ is related to $Z_{WRT}$ by the same formula as  $Z_{TV}$ to $Z_{WRT}$, which means that $\tilde{Z}_{TV}$ and $Z_{TV}$ coincide. One can also show this directly, by using the Chain-Mail techniques, and one obtains that $Z_{TV} = \tilde{Z}_{TV}$ \cite{M1}.

\item In the case $M= \Sigma \times I$, where $I=[0,1]$, $\tilde{Z}_{TV}$ coincides with the spin foam state-sum invariants constructed in \cite{M3} for spin networks embedded in $\Sigma$. In particular, when $\Sigma = S^2$ then $\tilde Z_{TV} (\Gamma,S^2 \times I)$ is proportional to the quantum group evaluation of the spin network $\Gamma$, which means that the evaluation of a spin network can be expressed as a sum of products of $6j$ symbols. This can be used for obtaining the large-spin asymptotics of the quantum group evaluation of a spin network, since the large-spin asymptotics of the quantum $6j$ symbols is known \cite{MT}.

\end{itemize}
 
\section{Conclusions}

The spin foam state sum invariants of 3-manifolds and embedded graphs can be constructed by using the essentially finite category of representations of the quantum group $U_q (\bf g)$ for $q$ a root of unity. In the $SU(2)$ group case these state sums can be identified with the path integrals for the 3d Euclidean GR with a non-zero cosmological constant $\lambda$. An interesting question is how to define these state sums when $\lambda =0$, since then the finite category of the quantum group representations is replaced by the infinite category of the Lie group representations. It turns out that it is possible to define the path integral in this case, by regularising the group integral (\ref{dpi}) \cite{BG}. A finite invariant can be obtained only in special cases, when the second twisted cohomology group vanishes. This is in agreement with the results of Witten \cite{W2}, who showed that the $\lambda =0$ path integral can be represented as an integral over the space of flat connections of the Reidemeister torsion, and the integral exists if the $B$ field does not have certain zero-modes.

Note that we have discussed the spin-foam and related CS theory invariants for the case of a compact Lie group $G$. The case when $G$ is noncompact is an outstanding problem, and sveral strategies have been suggested so far:
\begin{itemize}

\item Repeat the steps of sect 1. for the category of unitary irreps of $G$. Because the unitary irreps of non-compact Lie groups are infinite-dimensional, just defining the classical spin network evaluations is a non-trivial task. Then one has to regularize the resulting formal sums by finding a finite category of quantum group representations. However, nobody has succeded so far in finding an appropriate category of the quantum group irreps, see \cite{FM,BC}.

\item Try to regularize the group integral (\ref{dpi}), by using the approach of \cite{BG}.

\item Try to find an analytic continuation of the compact group results, see \cite{FM,W3}.
\end{itemize}

\bibliographystyle{amsalpha}

\end{document}